\documentclass[lettersize,journal]{IEEEtran}
\usepackage{cite}
\usepackage{amsmath,amssymb,amsfonts}
\usepackage{algorithmic}
\usepackage{algorithm}
\usepackage{graphicx}
\usepackage{algorithm,algorithmic}
\usepackage{hyperref}
\hypersetup{hidelinks=true}
\usepackage{textcomp}
\def\BibTeX{{\rm B\kern-.05em{\sc i\kern-.025em b}\kern-.08em
    T\kern-.1667em\lower.7ex\hbox{E}\kern-.125emX}}
\usepackage{balance}
\usepackage{subfig}
\usepackage{cite}
\usepackage{tikz}
\usepackage{tkz-tab}
\usepackage{pgfplots}
\usepackage{pgfplotstable}
\usepackage{caption}
\usepackage{amsmath}
\usepackage{longtable}
\usepackage{amssymb}
\usepackage{multicol}
\usepackage{multirow}
\usepackage{anyfontsize}

\begin{document}
\title{A Global Cybersecurity Standardization Framework for Healthcare Informatics}
\author{Kishu Gupta, \IEEEmembership{Member, IEEE,} Vinaytosh Mishra, and Aaisha Makkar, \IEEEmembership{Member, IEEE}
\thanks{Received 17 May 2024; revised 26 August 2024 and 18 September 2024; accepted 21 September 2024. This work is supported by National Sun Yat-sen University, Kaohsiung, Taiwan; Thumbay Institute for AI in Healthcare, Gulf Medical University, Ajman, UAE; University of Derby, UK. (Corresponding author: Vinaytosh Mishra.)}
\thanks{Kishu Gupta is with the Department of Computer Science and Engineering, National Sun Yat-sen University, Kaohsiung, 80424, Taiwan (e-mail: kishuguptares@gmail.com).}
\thanks{Vinaytosh Mishra is with the Thumbay Institute for AI in Healthcare, Gulf Medical University, Ajman, United Arab Emirates (e-mail: dr.vinaytosh@gmu.ac.ae).}
\thanks{Aaisha Makkar is with the College of Science \& Engineering, University of Derby, UK (e-mail: a.makkar@derby.ac.uk).}}

\makeatletter
\newcommand{\removelatexerror}{\let\@latex@error\@gobble}
\def\ps@IEEEtitlepagestyle{%
	\def\@oddfoot{\mycopyrightnotice}%
	\def\@oddhead{\hbox{}\@IEEEheaderstyle\leftmark\hfil\thepage}\relax
	\def\@evenhead{\@IEEEheaderstyle\thepage\hfil\leftmark\hbox{}}\relax
	\def\@evenfoot{}%
}

\def\mycopyrightnotice{%
	\begin{minipage}{\textwidth}
		\centering \scriptsize
		This article has been accepted for publication in IEEE Journal of Biomedical and Health Informatics. This is the author's version which has not been fully edited and content may change prior to final publication. Citation information: DOI 10.1109/JBHI.2024.3467179.\\This article has been accepted in IEEE Journal of Biomedical and Health Informatics © 2024 IEEE. Personal use of this material is permitted. Permission from IEEE must be obtained for all other uses, in any current or future media, including reprinting/republishing this material for advertising or promotional purposes, creating new collective works, for resale or redistribution to servers or lists, or reuse of any copyrighted component of this work in other works. This work is freely available for survey and citation.
		
	\end{minipage}
}
\makeatother

\maketitle

\begin{abstract}
Healthcare has witnessed an increased digitalization in the post-COVID world. Technologies such as the medical internet of things and wearable devices are generating a plethora of data available on the cloud anytime from anywhere. This data can be analyzed using advanced artificial intelligence techniques for diagnosis, prognosis, or even treatment of disease. This advancement comes with a major risk to protecting and securing protected health information (PHI). The prevailing regulations for preserving PHI are neither comprehensive nor easy to implement. The study first identifies twenty activities crucial for privacy and security, then categorizes them into five homogeneous categories namely: $\complement_1$ (Policy and Compliance Management), $\complement_2$ (Employee Training and Awareness), $\complement_3$ (Data Protection and Privacy Control), $\complement_4$ (Monitoring and Response), and $\complement_5$ (Technology and Infrastructure Security) and prioritizes these categories to provide a framework for the implementation of privacy and security in a wise manner. The framework utilized the Delphi Method to identify activities, criteria for categorization, and prioritization. Categorization is based on the Density-Based Spatial Clustering of Applications with Noise (DBSCAN), and prioritization is performed using a Technique for Order of Preference by Similarity to the Ideal Solution (TOPSIS). The outcomes conclude that $\complement_3$ activities should be given first preference in implementation and followed by $\complement_1$  and $\complement_2$ activities. Finally, $\complement_4$ and $\complement_5$ should be implemented. The prioritized view of identified clustered healthcare activities related to security and privacy, are useful for healthcare policymakers and healthcare informatics professionals.
\end{abstract}


\section{Introduction}
\label{sec:introduction}
\IEEEPARstart{H}{ealthcare} has witnessed rapid digitalization in recent years \cite{c1article}. Industry 4.0 and its main enabling information and communication technologies are completely changing services and production. This is especially true for the health domain, where the Internet of Things, cloud, and big data technologies are revolutionizing eHealth and its whole ecosystem, moving it towards \cite{c2ACETO2020100129, MAIDS}. Technologies such as Artificial Intelligence (AI), Internet of Things (IoT), and Cloud Computing are transforming the way healthcare has been delivered in the recent past \cite{c3MISHRA2022100684, FedMUPGUPTA2024111519}. AI enables systems to augment medical staff in each aspect of care, including diagnosis, prognosis, and treatment. These technologies impact the efficiency of nursing and the managerial activities of hospitals \cite{c2ACETO2020100129, c4ijerph18010271, Forecast10.1007/978-981-15-8335-3_1}. Healthcare is one of the most prominent fields utilizing IoT. This technology enables medical practitioners and hospital staff to perform their duties efficiently and intelligently. With the latest advanced technologies, most of the challenges of using IoT have been resolved, and this technology can be a great revolution and has many benefits in the future of digital healthcare \cite{c2ACETO2020100129, c5AGHDAM2021105903, INFOCOMP}. Similarly, cloud computing has transformed the traditional way of healthcare delivery \cite{GDSPS10461067}. It offers numerous healthcare advantages, significantly impacting how healthcare data is managed, accessed, and leveraged for patient care \cite{c2ACETO2020100129}. Cloud-based Electronic Health Records (EHRs) are key enablers of digital health. It facilitates the digital retrieval of patient records and the extraction of clinical information. Consequently, various additional uses of this technology have become available, including quality management, healthcare administration, and translational research \cite{c7Friedman2013ConceptualisingAC}. 

Chenthara et al. discuss the risks to privacy and security in cloud-based healthcare records, suggesting mitigation strategies \cite{c88726303}. They suggest pseudonymizing Electronic Health Record (EHR) data to safeguard Personal Health Information (PHI) privacy and security. 
Recent research, including multiple studies, points to blockchain technology as a promising solution for privacy and security issues in EHRs \cite{c10article, c118863359}. Tang et al. describe an efficient blockchain-based scheme \cite{c12article}, and Guo et al. offer a secure, attribute-based signature scheme for blockchain in EHRs involving multiple authorities \cite{c13article}. Al Mamun et al. thoroughly examine blockchain's application in EHRs and outline areas for future research \cite{c14article}. 


Ensuring privacy and security necessitates a comprehensive overhaul of infrastructure, processes, and practices within healthcare organizations, alongside allocating necessary resources for effective implementation. Moreover, prioritizing these efforts helps implement these measures in resource-constrained settings. Based on the review of the extant literature, we observe there is a lack of a comprehensive healthcare-specific privacy and security framework. While HIPAA is widely accepted, it falls short in addressing challenges posed by emerging technologies like tele-health, AI, and cross-border data sharing. A high need emerges to provide a technical solution that is sufficient; considering behavioral and awareness factors are equally crucial in managing PHI's privacy and security in healthcare.

\textit{Paper Contributions and Outline: } This study aims to develop a tailored framework to enhance data protection in modern healthcare systems by proposing a \textbf{G}lobal \textbf{C}yber-\textbf{S}ecurity \textbf{S}tandardization Framework for \textbf{H}ealthcare \textbf{I}nformatics (GCS-HI). 
The key contributions of the outlined model are as follows:
\begin{enumerate}
\item The study proposes a novel, three-fold structured GCS-HI framework to enhance data privacy and security in the healthcare system. It strategically identifies, categorizes, and prioritizes the activities crucial towards privacy and security using brain-storming for identification, clustering for categorization, and the multi-criteria decision-making (MCDM) approach for prioritization.
\item The framework offers a systematic method for healthcare organizations to adopt comprehensive security measures in a post-COVID digitalized environment.
\item The experimental work and extensive comparison with state-of-the-art approaches highlight the importance and efficiency of the proposed GCS-HI framework. 
\end{enumerate}
The roadmap of the paper is as follows: Section \ref{sec:rel} outlines the related literature for the study. Section \ref{sec:proposed} describes the proposed model framework. Followed by Section \ref{sec:res} that discusses the results. Finally, Section \ref{sec:con} concludes with a summary, research implications, and future direction for research.  

\section{Related Work} \label{sec:rel}
The terms "privacy," "confidentiality," and "security" in the context of healthcare are interrelated but distinct concepts. The privacy means that a patient's personal health information is only accessible to the patient and those authorized by the patient \cite{c16HATHALIYA2020311, c179484786}. Meanwhile, confidentiality obligates healthcare providers to protect patient information and disclose it only with the patient's consent or under legally permissible circumstances \cite{c179484786, c18article}. Security refers to the measures, protocols, and procedures that protect personal or sensitive information from unauthorized access, disclosure, alteration, and destruction \cite{c16HATHALIYA2020311}. The literature study is divided into two major aspects; first, the threats in healthcare related to privacy and security, and second, the widely accepted healthcare standards standards in practice. 
\subsection{Privacy and Security Threats}
Due to the sensitive nature of data, privacy and security threats are a major concern in healthcare. Some of the major threats reported in the extant literature are: 
\subsubsection{Data Breaches and Cyberattacks} 
Cybercriminals attack health information systems to steal PHI, which is in high demand in the market \cite{c19Wasserman2022HospitalCR}. One of the common methods is an attack of ransomware, where attackers encrypt data and demand payment for its release \cite{c20article}. 
\subsubsection{Insider Threats}
Healthcare staff can misuse their access to PHI, which can result in privacy breaches. They can do it either maliciously or accidentally. This can include viewing patient records without a legitimate reason or inadvertently disclosing information \cite{c21Saxena2020ImpactAK}. Prabhu \& Thompson propose a unified classification model of insider threats to information security \cite{c22article}.
\subsubsection{Phishing and Social Engineering Attacks}
Phishing attacks intend to trick healthcare employees into revealing sensitive information, such as login credentials. It is a major privacy invasion in which an attacker poses as a legitimate entity to gain access \cite{c23info13080392}. The approach, such as social engineering, is popular among hackers with malicious intentions \cite{c249743471}. 
\subsubsection{Inadequate Security Measures and Policies}
Healthcare providers can fail to implement sufficient security measures such as data encryption or access controls \cite{c25article}. Lack of planned security audits and training can expose an organization to privacy and confidentiality risks \cite{c26SAROSH2021100225}. 
\subsubsection{Mobile Device Security}
Ubiquitous personal devices and hospital EHRs on these systems expose a health system to privacy and security risks \cite{c279130678}. The increasing use of mobile devices in healthcare, such as tablets and smartphones, can create security vulnerabilities, especially if these devices are lost, stolen, or used over unsecured networks \cite{c28article}.
\subsubsection{Unsecured IoT Devices}
Medical IoT devices are designed foremost for usability, but with this simplicity of design, most fail to support encryption \cite{c29article}. This means that whenever a medical IoT device is used to connect with a hospital network or healthcare database, there is a risk of interception or infiltration \cite{c30su132111645}. 
\subsubsection{Lack of Patient Awareness}
Actions like sharing personal health information on unsecured platforms or falling for scams result in the breach of privacy and security of patients \cite{c31article}. A need for patient education program is required to mitigate the risk of security breaches originating because of the patient's irresponsible behavior \cite{c18article}. 
\subsection{Privacy and Security Standards and Law}
The most widely used laws for ensuring privacy and security include the Health Insurance Portability and Accountability Act (HIPAA) from the US and the General Data Protection Regulation (GDPR) from the European Union \cite{c32}. Other major laws include the Personal Information Protection and Electronic Documents Act (PIPEDA) of Canada, the Health Records and Information Privacy Act (HRIPA) of Australia, the Data Protection Act (DPA) of the UK, and Digital Personal Data Protection Act (DPDPA) of India. Most of these laws are implementations of GDPR with provisions of HIPAA in healthcare contexts. These laws have evolved and have been mostly reactive towards the privacy and security threats \cite{c33Moore17}. Moreover, the continually evolving state-of-the-art techniques in Machine Learning (ML), Data Analytics (DA), and hacking are making it even more difficult to protect a patient's privacy absolutely \cite{c349146114}. 
\section{Proposed Framework} \label{sec:proposed}
\begin{figure}[!htbp]
	\centering
	\includegraphics[width=1.0\columnwidth]{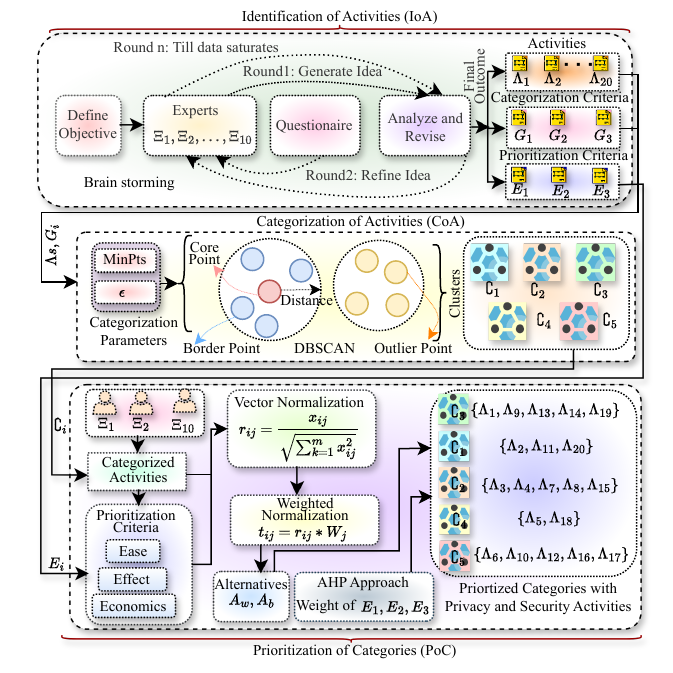}
	\caption{GCS-HI schematic overview.}
	\label{fig:proposed}
\end{figure}
The GCS-HI framework considers the condition of various worldwide healthcare standards ($\hslash$) such as: \{HIPAA, GDPR, PIPEDA, HRIPA, DPDPA, GDPR\} comprising considerable activities: $\{\Lambda_1, \Lambda_2, \ldots, \Lambda_n\} \in \Lambda$ crucial towards privacy and security of healthcare data as illustrated in Fig. \ref{fig:proposed}. A lot of healthcare privacy and security-related activities ($\Lambda$) are common in all $\hslash$ while each ($\hslash$) has a few unique activities as well. The proposed framework identifies the most pivotal $\Lambda$s among these and the categorization criteria $\{G_1, G_2, G_3\}$ for the $\Lambda$s by performing brainstorming to utilize the expertise of a focus group of ten experts: $\{\Xi_1, \Xi_2, \ldots, \Xi_{10}\} \in \Xi$ having different educational backgrounds with diverse professions. Further, the categorization parameters such as $\epsilon$ and MinPts are employed to categorize $\Lambda$s into different categories: $\{\complement_1, \complement_2, \ldots, \complement_5\} \in \complement$. These $\Xi$s prioritizes the $\complement$s by utilizing the three identified prioritization criteria: $\{E_1, E_2, E_3\}$ means the \{Ease, Effect, Economics\} on a weighted scale of [1,10] for each $E$. The detailed description regarding the identification of activities (IoA), categorization of activities (CoA), and prioritization of categories (PoC) are described in the following subsections: \ref{subsec:pro1}, \ref{subsec:pro2} and, \ref{subsec:pro3}, respectively. 
\subsection{Identification of Activities (IoA)} \label{subsec:pro1}
To identify the key activities ($\Lambda$s), the proposed framework utilizes the brainstorming-based-delphi approach. The existing literature recommends a considerable size of eight to sixteen experts ($\Xi$). It selects the participants $\Xi$ using purposive sampling and thus considers a focus group of ten $\Xi$s to perform the brainstorming activities. The delphi study involved three iterative rounds: identifying activities, establishing categorization criteria, and prioritizing them. In cases of indecision, a simple majority was used to reach a decision. This process ensured a thorough, expert-driven consensus, producing a well-structured and prioritized set of activities and criteria. Table \ref{tab:expert} elaborate the details of considered focus group of $\Xi$s 
\begin{table}[!ht]
	\caption{Characteristics of focus group of experts ($\Xi$)}
	\label{tab:expert}
	\centering
	\resizebox{0.9\columnwidth}{!}{
		\tiny
		\begin{tabular}{|l|l|l|c|}
		\hline
		\textbf{Expert ($\Xi$)} & \textbf{Profession} & \textbf{Education} & \textbf{Experience (years)} \\
		\hline
				$\Xi_1$ & SE & B.Tech  & 5 \\ 	\hline
				$\Xi_2$ & SE & B.Tech & 6 \\ 	\hline
				$\Xi_3$ & SE & M.Tech  & 5 \\ 	\hline
				$\Xi_4$ & SE & MCA  & 6 \\ 	\hline
				$\Xi_5$ & WA  & MCA  & 8 \\ 	\hline
				$\Xi_6$ & DA & B.Tech & 6 \\ 	\hline
				$\Xi_7$ & HM  & B.Tech  & 6 \\ 	\hline
				$\Xi_8$ & HM  & MBA   & 8 \\ 	\hline
				$\Xi_9$ & HM  & MBA   & 9 \\ 	\hline
				$\Xi_{10}$ & HM  & MBA  & 8 \\ \hline \hline
				\noalign{\smallskip} 
	\end{tabular}}

	\footnotesize{SE: Software Engineer; WA: Web Analyst; DA: Data Analyst; HM: Healthcare Manager;  B.Tech: Bachelor of Technology; M.Tech: Master of Technology; MCA: Master of Computer Applications; MBA: Master of Business Administration}
\end{table}

The brainstorming process as illustrated in Fig. \ref{fig:proposed} comprises major steps like defining the objective, selecting of a focus group of $\Xi$s, multiple rounds of analysis till data saturates, and lastly utilizing these results for identification of $\Lambda_i: \{\Lambda_1, \Lambda_2, \ldots, \Lambda_n\}$, $G_i: \{G_1, G_2, G_3\}$, and $E_i: \{E_1, E_2, E_3\}$. Before initiating the brainstorming session, a focus group of participants ($\Xi$s) are briefed on diverse global medical standards ($\hslash$) such as HIPAA and GDPR. During the refinement phase, any recurring themes were identified and eliminated. A voting method is conducted in case a clear decision can not be made. The iteration required for the identification of $\Lambda_i$, $G_i$, and $E_i$  were 5, 3, and 2, respectively. 
\subsection{Categorization of Activities (CoA)} \label{subsec:pro2}
\begin{figure}[!ht]
	\centering
	\includegraphics[width=0.6\columnwidth,height=3.0cm]{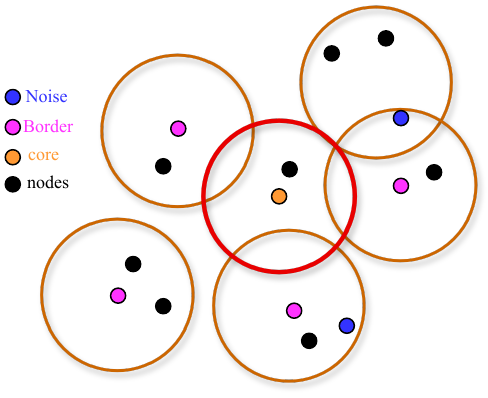}
	\caption{Illustration of density based categorization approach.}
	\label{fig2}
\end{figure}

To categorize the identified activities $\{\Lambda_1, \Lambda_2, \ldots, \Lambda_n\}$ into effectively manageable groups, the proposed GCS-HI framework deployed a density-based spatial clustering of applications with noise (DBSCAN) clustering algorithm because of its capability in handling the noise, cluster shape flexibility, proficiency in automatically detecting clusters, resilience against outliers, capability to manage clusters of irregular shapes, effectiveness in identifying noise, adaptability to different data types and dimensions, efficiency, scalability, and ease of interpretation. The clustering process involved two key parameters: $\epsilon$ and MinPts, where $\epsilon$ is the distance threshold that defines the neighborhood around a data point. At the same time, MinPts is the minimum number of points required to form a dense region. A typical depiction of the clustering approach is given in Fig. \ref{fig2}. 

It assumes, $D$ is the dataset containing $n$ points: $D= \{p_1, p_2, ..., p_n\}$.  Consider all points $P_i$ in the dataset $D$ are visited.  All the border points in $\epsilon$-Neighborhood of point $P_i$ ($P_i : N_{\epsilon}(P_i)$) are computed by using Eq. (\ref{eq:b}) and all $P_i$ are added to the new cluster $\complement_k$. Further, the $\complement_k$ is expanded for un-visited points $P_j$ in $N_{\epsilon}(P_i)$ and to mark $P_j$ as visited. Now, to find all points in $\epsilon$-Neighborhood of point $P_j$ by using Eq. (\ref{eq:c}) and these points are added to  $N_{\epsilon}(P_i)$ for further processing and if$P_j$ is not yet a member of any cluster add $P_j$ to $\complement_k$. 
\begin{gather}
	\label{eq:b}
	|N_{\epsilon}(P_i)|  < MinPts \\
	\label{eq:c}
	N_{\epsilon}(P_j) if |N_{\epsilon}(P_j)|  \geq MinPt
\end{gather}

For any possible clustering $\{\complement_1, \complement_2, \ldots, \complement_5\}$ out of all clusters $\complement$, it minimizes the number of clusters under the condition that every pair of points in a cluster is densely reachable, which corresponds to two main properties, connectivity, and maximality of the cluster. The main objective is to optimize the loss function as computed using Eq. (\ref{eq:loss}). 
\begin{align} 	\label{eq:loss}
	\begin{split}
		\min \vert \complement \vert \\
		\complement \subset \complement \\
		d_{dc}(p,q) \leq \epsilon, \forall \, p,q \in \complement_i \forall \complement_i \in \complement 
	\end{split}
\end{align}
wherein, $d_{db}(p,q)$, gives the smallest $\epsilon$ such that two points $p$ and $q$ are density-connected.

The distance function is computed using Eq. (\ref{eq1}) denotes the distance between two points, $p$ and $q$ in m-dimensional space. The neighborhood function is computed using Eq. (\ref{eq2}) denotes $\epsilon$-neighborhood of point $p$. Algorithm \ref{alg:DBSCAN} presents a summarized form of the approach used for categorization. 
\begin{gather}
	\label{eq1}
	dist(p,q)= \sqrt[2]{\sum_{i=1}^{m}(p_i-q_i)^2} \\ 
	\label{eq2}
	N_{\epsilon}(p) = \{q\in D \;|\; dist(p,q)\leq \epsilon \}
\end{gather}
\subsection{Prioritization of Categories (PoC)} \label{subsec:pro3}
Followed by the classification of the most admissible categories ($\complement$)s, the prioritization of these categories is performed. To perform the prioritization a multi-criteria decision-making approach called the Technique for Order of Preference by Similarity to the Ideal Solution (TOPSIS) is utilized. TOPSIS is used for multi-criteria decision-making because it provides a clear, systematic method for ranking alternatives based on their relative closeness to an ideal solution. TOPSIS is a robust and reliable tool in scenarios requiring the careful balancing of various factors. The given decision data is normalized by deploying the vector normalization approach using Eq. (\ref{eq3}). The normalization matrix represents the normalization of evaluation matrix comprising $n$ alternatives activities ($\Lambda$) and $m$ criteria ($\mathfrak{C}$) such that $(x_{ij})_{m*n}$. 
\begin{equation}
	\label{eq3}
	r_{ij} = \frac{x_{ij}}{\sqrt{\sum_{k=1}^{m} x_{ij}^2}}
\end{equation}
here, $i = 1, 2,..., m$ and $j = 1, 2,..., n$. 
The weighted normalized decision matrix ($t_{ij} $) is computed using Eq. (\ref{eq4}) and the weight of criteria for prioritization is computed using the analytical hierarchy process (AHP) Approach. 
\begin{equation}
	\label{eq4}
	t_{ij} = r_{ij}*W_j 
	\end{equation}
Further, the worst ($A_w$) and best alternatives ($A_b$) are computed using Eq. (\ref{eq5}). Moreover, Eqs. (\ref{eq6}) and (\ref{eq7}), computes the distance between target $i$ and worst or bad condition, respectively. 
\begin{algorithm}[H]
	\caption{CoA: Summary for Categorization.}\label{alg:DBSCAN}
	\begin{algorithmic}
		\STATE 
		\STATE Label all points as un-visited
		\FOR{each point $P_i$ in the dataset $D$}
		\IF{$P_i$ is visited}
		\STATE Skip to next point
		\ENDIF	
		\STATE Mark $P_i$ as visited
		\STATE find all points in $\epsilon$-Neighborhood of point $P_i$. Such that $P_i : N_{\epsilon}(P_i)$
		\IF{$|N_{\epsilon}(P_i)|  <$ MinPts}
		\STATE Mark $P_i$ as noise (later it may be classified as border point)
		\ELSE
		\STATE create a new cluster, $\complement_k$ and add $P_i$ to $\complement_k$ 	
		\ENDIF	
		\STATE Expand Cluster $\complement_K$
		\FOR{each point $P_j$ in $N_{\epsilon}(P_i)$}
		\IF{$P_j$ is unvisited}
		\STATE Mark $P_j$ as visited
		\ENDIF	
		\STATE find all points in $\epsilon$-Neighborhood of point $P_j$. Such that $P_j : N_{\epsilon}(P_j)$
		\IF{$|N_{\epsilon}(P_j)|  \geq$ MinPts}
		\STATE add these points to  $N_{\epsilon}(P_i)$ for further processing (expand the neighborhood)
		\ENDIF
		\ENDFOR	
		\IF{$P_j$ is not yet a member of any cluster}
		\STATE add $P_j$ to $\complement_k$
		\ENDIF	
		\ENDFOR	
	\end{algorithmic}
\end{algorithm}

\begin{gather} 
	\label{eq5}
		\begin{split}
			A_w = \biggl\{\langle max (t_{ij}\bigg| i=1, 2, ..., m) \ j \epsilon J_- \rangle, \\
			\langle min (t_{ij}\bigg| i=1, 2, ..., m) \ j \epsilon J_+ \rangle \biggr\} \\
			 \equiv \{t_{wj}| j =1, 2, ..., n\} \\
			A_b = \biggl\{ \langle min (t_{ij}\bigg| i=1, 2, ..., m) \ j \epsilon J_- \rangle, \\
			\langle max (t_{ij}\bigg| i=1, 2, ..., m) \ j \epsilon J_+ \rangle \biggr\} \\
			\equiv \{t_{bj}| j =1, 2, ..., n\}
		\end{split}
\end{gather}
wherein, $J_+ = \{j = 1, 2, ..., n| j\}$ and $J_- = \{j = 1, 2, ..., n| j\}$ are associated with a criteria having positive impact and negative impact, respectively.
\begin{gather}
	\label{eq6}
	d_{iw} = \sqrt{\sum_{j=1}^{n} (t_{ij}-t_{wj})^2} \\
	\label{eq7}
		d_{ib} = \sqrt{\sum_{j=1}^{n} (t_{ij}-t_{bj})^2}
\end{gather}
here, $i = 1, 2, ..., m$. 

The effectiveness of prioritization depends on the proper identification of criteria and accurate weighting. The steps used for the prioritization process are demonstrated in Algorithm \ref{alg:TOPSIS}. The prioritization using TOPSIS has several advantages over other multi-criteria decision-making such as simplicity and understand ability, balance between ideal and negative ideal solutions, clear ranking of alternatives, and compatibility with other methods. 
\begin{algorithm}[H]
	\caption{PoC: Summary for Prioritization.}\label{alg:TOPSIS}
	\begin{algorithmic}
		\STATE 
		\STATE \textbf{Start}
		\STATE Create an evaluation matrix consisting of $n$ activities and $n$ criteria $(x_{ij})_{m*n}$
		\STATE The matrix $(x_{ij})_{m*n}$ is normalized to obtain $R = (r_{ij})_{m*n}$ as shown in Eq. (\ref{eq3})
		\STATE Calculate the weighted normalized decision matrix ($t_{ij}$) using Eq. (\ref{eq4})
		\STATE Determine the worst alternative $A_w$ and best alternative $A_b$ using Eq. (\ref{eq5})
		\STATE Calculate the distance between target alternative $i$ and worst condition $A_w$ using Eq. (\ref{eq6})  
		\STATE Similarly, calculate the distance between target alternative $i$ and best condition $A_b$ using Eq. (\ref{eq7}) 
		\STATE \textbf{End}
	\end{algorithmic}
\end{algorithm}
\section{Operational Design and Complexity}
The overall summary of the proposed GCS-HI framework is illustrated in Algorithm \ref{alg:GCS-HI}. The study used a naive implementation for the categorization of activities without any indexing or optimized search for the neighborhood. 
\begin{algorithm}[H]
	\caption{GCS-HI: Operational Summary.}\label{alg:GCS-HI}
	\begin{algorithmic}
		\STATE 
		\STATE \textbf{Start}
		\STATE Define the objective
		\STATE Select Experts : $\{\Xi_1, \Xi_2, \ldots, \Xi_{10}\}$
		\STATE Perform IoA by deploying brain-storming approach to identify the pivotal activities: $\Lambda_1, \Lambda_2, \ldots, \Lambda_n$
		\STATE Perform CoA by using Eqs. (\ref{eq:b})-(\ref{eq2}) as described in Algorithm \ref{alg:DBSCAN}
		\STATE Perform PoC by using  Eqs. (\ref{eq3})-(\ref{eq7}) as explained in Algorithm \ref{alg:TOPSIS}
		\STATE Output: Pivotal activities in the order of their importance
		\STATE \textbf{End}
	\end{algorithmic}
\end{algorithm}

\textit{Complexity:} The time complexity is $O(n^2)$, where $n$ is the number of points. This is because, for each of the $n$ points, the algorithm needs to compute the distance to every other point to determine if they fall within the specified $\epsilon$-neighborhood. The algorithmic complexity of prioritization depends on the number of alternatives ($n$) and the number of criteria ($m$) involved in the decision-making process. The method consists of several computational steps, including normalization, weighting, determining the positive and negative ideal solutions, calculating distances, and finally, computing the similarity to the ideal solution. All steps have complexity $O(mn)$, while the similarity calculation step has complexity $O(m)$. The overall complexity of prioritization is thus $O(mn)$, as each step must be performed for every element in the decision matrix, and the steps are generally sequential. Therefore, the total time complexity of the GCS-HI framework is $\mathcal{O}(mn^3)$. 
\section{Performance Evaluation}\label{sec:res}
\subsection{Experimental Setup}
The experiment was carried out on machines with 11th Gen Intel\textsuperscript{\textregistered} Core \textsuperscript{TM} i7-1195G7 CPU, which employs a clock speed of 2.92 GHz. This computational system utilizes a 64-bit Windows 11 Home 22H2 version Operating system. The installed RAM of the system is 32.0 GB (31.8 GB usable). The software used for the DBSCAN was IBM SPSS 26.
\subsection{Description of Data}
The focus group was asked to rate identified activities based on the criteria identified. The activities are rated on criteria using an ordinal scale of [1 to 10], here 1 is the least, and 10 is the highest. The response average was taken and approximated to get the response matrix. The response matrix data was used for the categorization of activities using DBSCAN. Finally, categories were ranked using criteria of prioritization. The raw data used in the study was uploaded to an online repository, and DOI was generated \cite{c40}. 
\subsection{Results}
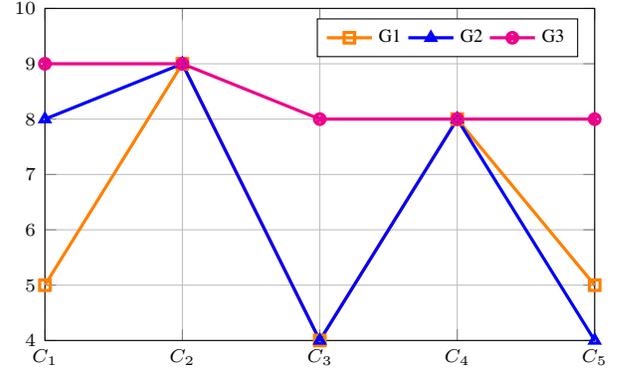
\begin{figure}[!htbp]
	\centering
		\begin{tikzpicture}
			\begin{axis}[
				xmin=1, xmax=5,
				ymin=4, ymax=10,
				xtick={1,2,3,4,5},xticklabels={$C_1$,$C_2$,$C_3$,$C_4$,$C_5$},
				xticklabel style = {font=\scriptsize},
				ytick={4,5,6,7,8,9,10},
				yticklabel style = {font=\scriptsize},
				legend pos=north east,
				legend style={font=\scriptsize, ,legend columns=3},
				ymajorgrids=true,xmajorgrids=true,
				height=6cm,
				width=3.5in
				]
				\addplot[color=orange,mark=square, very thick]
				coordinates {(1,5)(2,9)(3,4)(4,8)(5,5)};
				\addlegendentry{G1}
				\addplot[color=blue,mark=triangle, very thick]
				coordinates {(1,8)(2,9)(3,4)(4,8)(5,4)};
				\addlegendentry{G2}
				\addplot[color=magenta,mark=otimes, very thick]
				coordinates {(1,9)(2,9)(3,8)(4,8)(5,8)};
				\addlegendentry{G3}
			\end{axis}
		\end{tikzpicture} 
	\caption{Profile plot of the identified clusters.} \label{fig:r1}
\end{figure}

Based on brainstorming sessions, experts identified twenty activities required to preserve data privacy and security in healthcare. The focus group is asked to come up with twenty activities. The objective is achieved after five iterations. Secondly, the group is asked to identify criteria, for grouping these activities in meaningful ways. After three iterations, the focus group agreed on three criteria: Functional Focus (G1), Stakeholder Engagement (G2), and Strategic Objective (G3). The number of criteria was limited to three because of the inability of the DBSCAN method to handle multidimensional data. These twenty activities with their description are listed in Table \ref{tab2}. 
\begin{table}[!htbp]
	\caption{Pivotal activities to ensure privacy and security}
	\setlength{\tabcolsep}{3pt}
	\begin{tabular}{|p{25pt}|p{75pt}|p{130pt}|}
		\hline
		\textbf{Code} & \textbf{Activity ($\Lambda$)} & \textbf{Description of the Activity}  \\ \hline
		$\Lambda_1$ & Regular Risk Assessment & Conduct frequent risk assessments to identify potential vulnerabilities in the healthcare system. \\ \hline
		$\Lambda_2$ & Employee Training & Provide continuous training for employees on data privacy and security protocols. \\ \hline
		$\Lambda_3$ & Strong Access Controls & Implement strict access controls to ensure only authorized personnel can access sensitive data. \\ \hline
		$\Lambda_4$ & Data Encryption & Encrypt patient data in transit and at rest to protect against unauthorized access. \\ \hline
		$\Lambda_5$ & Audit Trails & Maintain detailed audit trails to monitor access and changes to patient data. \\ \hline
		$\Lambda_6$ & Anti-Malware Software & Install and regularly update anti-malware software to protect against cyber threats. \\ \hline
		$\Lambda_7$ & Secure Data Storage & Use secure storage solutions, such as encrypted databases, for patient data. \\ \hline
		$\Lambda_8$ & Data Minimization & Only collect and retain the minimum amount of patient data necessary for healthcare purposes. \\ \hline
		$\Lambda_9$ & Incident Response Plan & Develop and regularly update an incident response plan for potential data breaches. \\ \hline
		$\Lambda_{10}$ & Regular Software Updates & Keep all software and systems updated to protect against vulnerabilities. \\ \hline
		$\Lambda_{11}$ & Multi-Factor Authentication & Implement multi-factor authentication for accessing patient data systems. \\ \hline
		$\Lambda_{12}$ & Secure Communication Channels & Use secure communication channels, such as encrypted email, for transmitting patient data. \\ \hline
		$\Lambda_{13}$ & Patient Consent Management & Regularly obtain and manage patient consent for data use and sharing. \\ \hline
		$\Lambda_{14}$ & Third-Party Vendor Assessment & Conduct thorough assessments of third-party vendors who have access to patient data. \\ \hline
		$\Lambda_{15}$ & Data Anonymization Techniques & Apply data anonymization techniques where appropriate for research and analysis. \\ \hline
		$\Lambda_{16}$ & Physical Security Measures & Enhance physical security measures to protect data storage and access areas. \\ \hline
		$\Lambda_{17}$ & Mobile Device Management & Implement policies for secure use of mobile devices in accessing patient data. \\ \hline
		$\Lambda_{18}$ & Cybersecurity Insurance & Consider obtaining cybersecurity insurance to mitigate financial risks associated with data breaches. \\ \hline
		$\Lambda_{19}$ & Regular Compliance Audits & Conduct audits to ensure ongoing compliance with healthcare data protection regulations. \\ \hline
		$\Lambda_{20}$ & Patient Education & Educate patients about their data rights and how to protect their health information. \\
		\hline
	\end{tabular}
	\label{tab2}
\end{table}
The value of two parameters, $\epsilon$ and MinPts, is adjusted to zero noise points. Finally, a $\epsilon$ = 0.5 and MinPts = 2 are taken for the clustering assignment. The result of clustering is given in Table \ref{tab3}. The clusters are further given descriptive names for further analysis. 
\begin{table}[!htbp]
	\caption{Summary of categorization results}
	\setlength{\tabcolsep}{3pt}
	\begin{tabular}{|p{30pt}|p{90pt}|p{100pt}|}
		\hline
		\textbf{Cluster ($\complement$)} & \textbf{Member Activities} & \textbf{Name of Category} \\
		\hline
		$\complement_1$ & $\{\Lambda_1, \Lambda_9, \Lambda_{13}, \Lambda_{14}, \Lambda_{19}\}$ & Policy and Compliance Management  \\ \hline
		$\complement_2$& $\{\Lambda_2, \Lambda_{11}, \Lambda_{20}\}$ & Employee Training and Awareness \\ \hline
		$\complement_3$ & $\{\Lambda_3, \Lambda_4, \Lambda_7, \Lambda_8, \Lambda_{15}\}$ & Data Protection and Privacy Control  \\ \hline
		$\complement_4$ & $\{\Lambda_5, \Lambda_{18}\}$ & Monitoring and Response \\ \hline
		$\complement_5$ & $\{\Lambda_6, \Lambda_{10}, \Lambda_{12}, \Lambda_{16}, \Lambda_{17}\}$ & Technology and Infrastructure Security \\
		\hline
	\end{tabular}
	\label{tab3}
\end{table}
The profile plot for clusters using the mean value of criteria is calculated to see whether the category/cluster is distinct. Fig. \ref{fig:r1} shows that categories are different from each other. 

Further, the results of prioritizing categories using the TOPSIS method are explained. The three criteria taken for prioritization problems are (1) ease of implementation (Ease), (2) effectiveness (effect), and (3) economics (cost). All these criteria are beneficial as a high rating of economics means less cost of implementation. The weight of ease, effectiveness, and economy were calculated using AHP and found to be 0.11,0.63, and 0.26, respectively. The consistency ratio for pairwise comparison was 4\%, less than the recommended value of 10\%. Finally, the ranking categories identified are listed in Table \ref{tab3}. The initial decision matrix based on focus group discussion is given in Table \ref{tab4}. 
\begin{table}[!htbp]
	\caption{Final decision matrix for prioritization}
	\label{tab4}
		\centering
		\resizebox{1.0\columnwidth}{!}{
				\begin{tabular}{|p{30pt}|p{110pt}|p{15pt}|p{15pt}|p{35pt}|}
		\hline
		\textbf{Category ($\complement$)} & \textbf{Name of Category} & \textbf{Ease (E1)} & \textbf{Effect (E2)} & \textbf{Economics (E3)} \\
		\hline
		$\complement_1$ & Policy and Compliance Management  & 4 & 7 & 5 \\ \hline
		$\complement_2$ & Employee Training and Awareness & 4 & 8 & 6 \\ \hline
		$\complement_3$ & Data Protection and Privacy Control & 5 & 9 & 6 \\ \hline
		$\complement_4$	& Monitoring and Response & 6 & 7 & 6 \\ \hline
		$\complement_5$ & Technology and Infrastructure Security & 7 & 7 & 5 \\ 
		\hline
	\end{tabular}}
	
\end{table}

Now the decision matrix is normalized to obtain the matrix R, which is further multiplied with the criteria weight vector $W = [ 0.11, 0.63, 0.26 ]$ to get matrix T is given by: 
\[
T=
\begin{bmatrix}
	0.04 & 0.29 & 0.12  \\
	0.04 & 0.29 & 0.12  \\
	0.05 & 0.32 & 0.12  \\
	0.06 & 0.25 & 0.12  \\
	0.06 & 0.25 & 0.10 
\end{bmatrix}
\]
Based on the matrix T, the best alternative computed are represented as $A_b$ = [0.06, 0.32, 0.12] while the worst alternatives computed are given by $A_w$ = [0.04,0.25,0.10].
Now, the distance of alternatives (categories) from the best and worst alternatives is given by vectors $d_b$  =  [0.05, 0.05, 0.020, 0.07, 0.07] and $d_w $ = [0.04, 0.04, 0.08, 0.03, 0.03], respectively. Now, the vector representing similarity from the worst alternative $S_w$ = [0.48, 0.48, 0.80, 0.27, 0.27]. This helped to conclude that $\complement_3$ (data protection and privacy control) activities should be given first preference in implementation. $\complement_1$ and $\complement_2$ activities (policy and compliance management and employee training and awareness) should be implemented next. Finally, $\complement_4$ and $\complement_5$ (monitoring and response and technology and infrastructure security) should be implemented. The results of the prioritization show that all the identified categories have significant importance, and none of these categories can be left without implementation. 
\subsection{Comparison}
To further analyze the efficiency of the GCS-HI framework, it is deployed with comparable schemes including DTEM \cite{c3MISHRA2022100684}, Soni et al. \cite{c8DBLP:journals/tii/SoniPPI23}, Zahrani et al. \cite{Zahrani}, Ansari et al. \cite{ansari2020}, and Mishra et al. \cite{GDSPS10461067} as described in Table \ref{tab:soa}. The GCS-HI is a comprehensive framework that integrates the identification, categorization, and prioritization of activities to safeguard PHI. It identifies the key activities to ensure privacy and security in healthcare and provides a clear road map for implementing these measures in a resource-constrained environment. 
\begin{table*}[!htbp]
	\caption{GCS-HI framework v/s state-of-the-art approaches}\label{tab:soa}
	\centering
	\resizebox{0.99\textwidth}{!}{
		\tiny
		\begin{tabular}{|l|c|c|c|c|c|c|c|c|}
			\hline 
			\multirow{3}{*}{\textbf{Study}} & \multicolumn{5}{c|}{\textbf{Framework}}  & \multicolumn{2}{c|}{\textbf{Feature}} & \multirow{3}{*}{\textbf{Computational}}  \\ 
			\cline{2-6} \cline{7-8}& $\textbf{Factors}$ &  $\textbf{Prioritization}$ & \textbf{Relation} & \textbf{Suggested} &\textbf{Methodology} & \multirow{2}{*}{$\Xi$}& \multirow{2}{*}{$f$}&   \\ 
			& $\textbf{Selection}$ &  $\textbf{Method}$ & \textbf{Establishment} & \textbf{Framework} &  &&& \textbf{Complexity}  \\ \hline 
			DTEM \cite{c1MISHRA2022100684} & SLR & $\divideontimes$ & $MDS$& TTFA& Delphi &8&9& $\mathcal{O}(nr|\Xi|)$ \\  \hline 
			Soni et al. \cite{c8DBLP:journals/tii/SoniPPI23}  & SLR& $\divideontimes$ & $\divideontimes$ &Not Holistic & Experimental Design &5 &6 & $\mathcal{O}(n^2.m\log m)$  \\  \hline 
			Zahrani et al. \cite{Zahrani}  &  SLR& MCDM & TFN&Descriptive & Fuzzy, ANP, TOPSIS &6&13& $\mathcal{O}(n.m^2)$ \\  \hline 
			Ansari et al. \cite{ansari2020}  & SF & MCDM & ISO 27005& Prescriptive& Fuzzy, TOPSIS &25&7 & $\mathcal{O}(n^2.m)$\\  \hline  
			GDSPS \cite{GDSPS10461067} & RS & MCDM & $\times$& Descriptive& K-Means, OPA &7&20& $\mathcal{O}(nkm\log m)$ \\  \hline  
			\textbf{GCS-HI} & Delphi & Structural & $\times$& Prescriptive & DBSCAN, AHP-TOPSIS &10 &20 &$\mathcal{O}(mn^3)$ \\  \hline  \hline
	\end{tabular}} 
	
	\footnotesize{$\times$: Absent; $\surd$: Present; $\divideontimes$: Not applicable; $\Xi$: Experts; $f$: Factors; $SLR$: Systematic Literature Review; $SF$:Survey Form; $RS$: Review of Standards; $MCDM$: Multi Criteria Decision Making; $PRISMA$: Preferred Reporting Items for Systematic Reviews and Meta-Analyses; $OPA$: Ordinal Priority Approach; $DBSCAN$: Density-Based Spatial Clustering of Applications with Noise; $AHP$: Analytical Hierarchical Approach; $TOPSIS$: Technique for Order of Preference by Similarity to Ideal Solution; $MDS$: Multi-Dimensional Scaling; $TTFA$: Task Technology Fit Analysis; $TFN$: Triangular Fuzzy Number; $ANP$: Analytic Network Method; $n$: Number of Experts; $m$: Number of Criteria/Factors; $k$: Number of Clusters; $r$: Number of Rounds} 
\end{table*}  
\section{Conclusion}\label{sec:con}
This study developed a comprehensive, easy-to-implement framework for the phase-wise implementation of measures for ensuring privacy and security. It identified and categorized key activities essential for maintaining data privacy and security, ultimately prioritizing them based on ease of implementation, effectiveness, and economic feasibility. The application of DBSCAN effectively grouped activities into distinct clusters, while TOPSIS provided a clear ranking, emphasizing the importance of data protection and privacy control. The result conclude that $\complement_3$ activities should be given first preference in implementation. $\complement_1$ and $\complement_2$ activities should be implemented next. Finally, $\complement_4$ and $\complement_5$ should be implemented. Implementing privacy and security in healthcare faces challenges like balancing data access with protection, managing complex regulations (e.g., HIPAA, GDPR), and addressing technological vulnerabilities. Limited resources and varying organizational capabilities further complicate efforts, making it difficult to ensure consistent, comprehensive safeguards across diverse healthcare settings. 

Future research should focus on adapting and evolving this framework in response to emerging technologies and threats, ensuring it remains relevant and effective in a rapidly changing digital landscape. Additionally, further validation of the framework in real-world healthcare settings would be invaluable, contributing to its practical utility and effectiveness in safeguarding patient data. Also, the usability and interoperability of the healthcare information systems can be explored.



\section*{References}

\bibliographystyle{IEEEtran}
\bibliography{reference}

\begin{IEEEbiography}[{\includegraphics[width=1in,height=1.25in,clip,keepaspectratio]{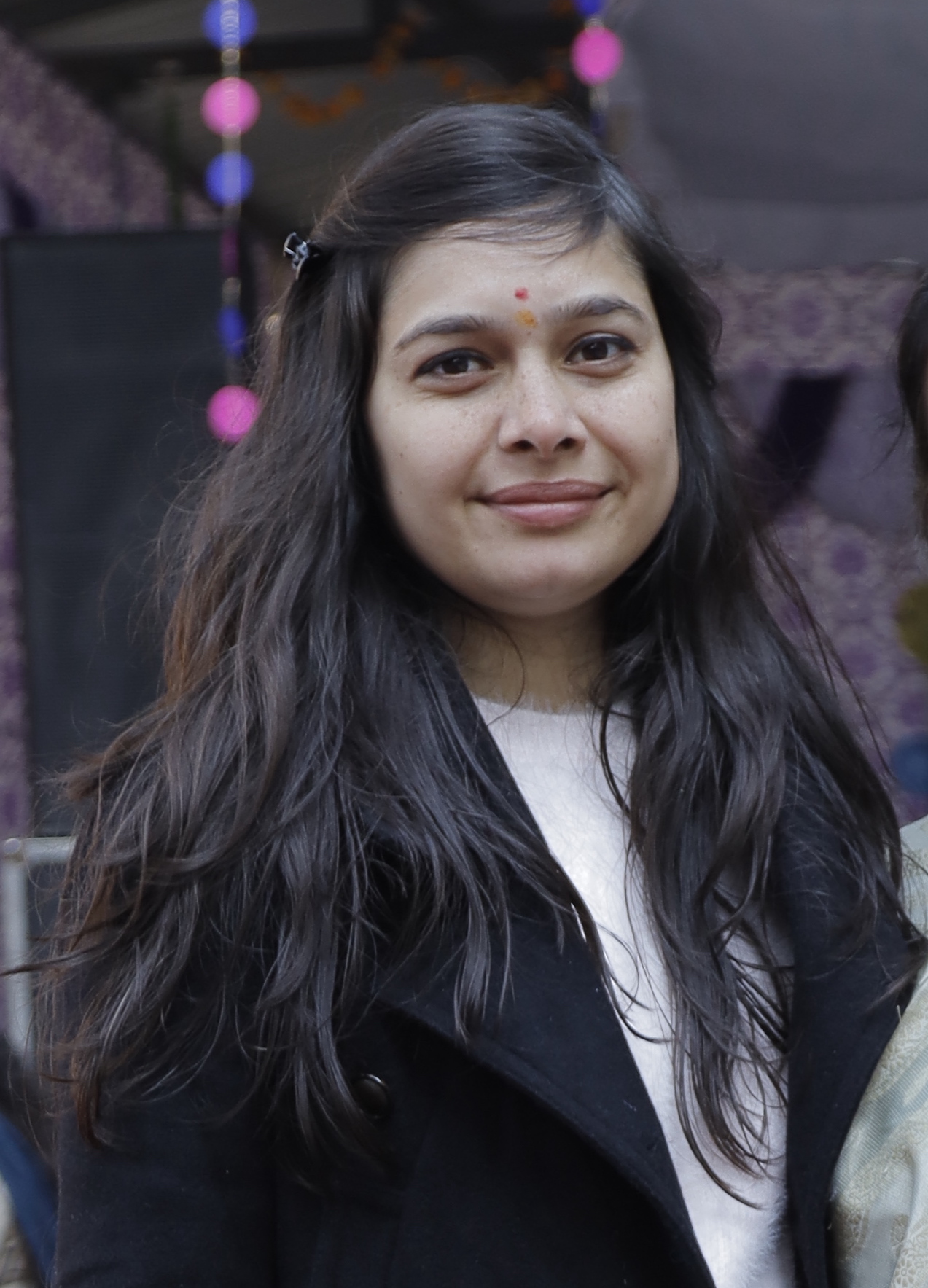}}]{Kishu Gupta} (Member, IEEE) is working as a Post Doctoral Research Fellow at the Cloud Computing Research Center, Department of Computer Science \& Engineering, National Sun Yat-sen University (NSYSU), Kaohsiung, Taiwan. She earned her Ph.D. from India in 2023 with prestigious \textit{INSPIRE Fellowship} sponsored by the Department of Science \& Technology (DST), under Ministry of Science and Technology (MOST), Govt. of India. Also, she is a recipient of the \textit{Gold Medal} for securing Ist rank in overall university during M.Sc. Her major research interest includes Data Security and Privacy, Cloud Computing, Federated Learning, Machine Learning, Quantum Computing, etc. Some of her research findings are published with top-notch venues including IEEE TASE, IEEE TCE, Applied Soft Computing, Cluster Computing, etc.
\end{IEEEbiography}
\vspace{11pt}
\begin{IEEEbiography}[{\includegraphics[width=1in,height=1.25in,clip,keepaspectratio]{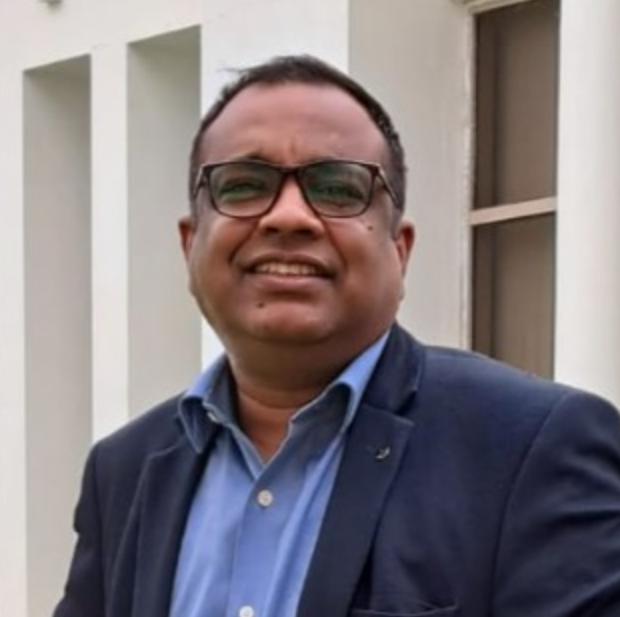}}]{Vinaytosh Mishra} is an Associate Professor and  Director of Thumbay Institute for AI in Healthcare, Gulf Medical University, Ajman, UAE. He earned PostDoc in AI in Healthcare from the University of Arizona, USA and PostDoc  in Ethical AI in Healthcare from the University of Ben Gurion, Israel. He has a PhD in Healthcare Management and a Bachelor of Technology in Electronics Engineering from the Indian Institute of Technology (BHU), India. He has over 19 years of experience in industries such as Information Technology, Manufacturing, Finance, Healthcare, and Education alongwith one Australian and One German Patent in AI in Healthcare. He has published more than 70 research papers in different journals and conferences of high repute. His current research interests include Statistics, Quality Assurance Engineering, Supply chain management, and Healthcare Digital Health. 
\end{IEEEbiography}
\vspace{11pt}

\begin{IEEEbiography}[{\includegraphics[width=1in,height=1.25in,clip,keepaspectratio]{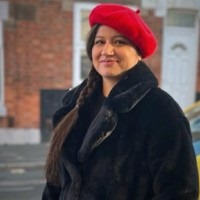}}]{Aaisha Makkar} (Member, IEEE)
 is a Lecturer in computer science at the University of Derby, UK. She is an experienced researcher with more than 8 years of cutting-edge research and teaching experience in prestigious higher education institutions, including University of Derby (UK), Seoul National University of Science and Technology (South Korea), and Thapar Institute of Engineering and Technology (India). She has authored and co-authored more than 40 research papers in high-ranked international journals (SCI indexed) and conferences. She has a track record of collaborations with industries, delivering innovative Artificial Intelligent based solutions to various emergent problems. 
\end{IEEEbiography}

\end{document}